\documentclass[journal,twocolumn,10pt,twoside]{IEEEtran}
\normalsize

\ifCLASSINFOpdf
   \usepackage[pdftex]{graphicx}
   \usepackage{epstopdf}
\else
\fi

\usepackage{amsmath}
\usepackage{amssymb}
\usepackage{cite}
\usepackage{graphicx}
\usepackage{algorithm}
\usepackage{algpseudocode}
\usepackage{subfigure}
\usepackage{multirow} 
\usepackage{longtable}
\usepackage{threeparttable}
\usepackage{caption3}
\usepackage{array}
\usepackage{cases}
\usepackage{color}
\usepackage[T1]{fontenc}
\usepackage[utf8]{inputenc}
\usepackage{authblk}

\newcommand\mycom[2]{\genfrac{}{}{0pt}{}{#1}{#2}}

\begin{document}
\title{Wirelessly Powered Backscatter Communications: Waveform Design and SNR-Energy Tradeoff}

\author{Bruno Clerckx, Zati Bayani Zawawi and Kaibin Huang  
\thanks{B. Clerckx and Z. Bayani Zawawi are with the EEE department at Imperial College London, London SW7 2AZ, UK (email: \{b.clerckx,z.zawawimohd-zawawi13\}@imperial.ac.uk). K. Huang is with the EEE department at University of Hong Kong (email: huangkb@eee.hku.hk). This work has been partially supported by the EPSRC of the UK under grant EP/P003885/1.
}
}

\maketitle

\begin{abstract} This paper shows that wirelessly powered backscatter communications is subject to a fundamental tradeoff between the harvested energy at the tag and the reliability of the backscatter communication, measured in terms of SNR at the reader. Assuming the RF transmit signal is a multisine waveform adaptive to the channel state information, we derive a systematic approach to optimize the transmit waveform weights (amplitudes and phases) in order to enlarge as much as possible the SNR-energy region. Performance evaluations confirm the significant benefits of using multiple frequency components in the adaptive transmit multisine waveform to exploit the nonlinearity of the rectifier and a frequency diversity gain. 
\end{abstract}


\IEEEpeerreviewmaketitle

\vspace{-0.3cm}
\section{Introduction}
\par The emergence of RFID technology in the last decade is the first sign of a serious interest for far-field wireless power transfer (WPT) and backscatter communications. RFID tags harvest energy from the transmit RF signal and rely on backscattering modulation to reflect and modulate the incoming RF signal for communication with an RFID reader. Since tags do not require oscillators to generate carrier signals, backscatter communications benefit from orders-of-magnitude lower power consumption than conventional radio communications \cite{Smith:2013}. Backscatter communication has recently received a renewed interest, in the context of the Internet-of-Things, with advances in backscatter communication theory and the development of sophisticated backscatter communication systems  \cite{Boyer:2014,Yang:2015,Kellogg:2016,Han:2017}. 
\par Backscatter communications commonly assume that the RF transmitter generates a sinusoidal continuous wave (CW). Significant progress has recently been made on the design of efficient signals for WPT \cite{Clerckx:2015,Clerckx:2016b,Huang:2016,Zeng:2017}. In particular, multisine waveforms adaptive to the Channel State Information (CSI) have been shown particularly powerful in exploiting the rectifier nonlinearity and the frequency-selectivity of the channel so as to maximize the amount of harvested DC power \cite{Clerckx:2016b}.
\par In this paper, we depart from this traditional CW transmission and leverage those recent progress in WPT signal design, and in particular the adaptive multisine wireless power waveform design, to show that wirelessly powered backscatter communications is subject to a fundamental tradeoff between the harvested energy at the tag and the SNR at the reader. Indeed, the SNR at the reader is a function of the backscatter channel (concatenation of the forward channel from transmitter to tag and backward channel from tag to reader) while the harvested energy at the tag is a function of the forward channel only. Due to the difference between those two channels, the optimal transmit waveform design for SNR and energy maximization are different. This suggests that adjusting the transmit waveform leads to a SNR-energy tradeoff.
\par Specifically, assuming that the CSI is perfectly available to the RF transmitter, we derive a systematic and optimal design of the transmit multisine waveform in order to enlarge as much as possible the SNR-energy region. Due to the non-linearity of the rectifier, the waveform design and the characterization of the region results from a non-convex posynomial maximization problem that can be solved iteratively using a successive convex approximation approach. Simulation results highlight that increasing the number of sinewaves in the transmit multisine waveform enlarges the SNR-energy region by exploiting the non-linearity of the rectifier and a frequency diversity gain.   
\par \textit{Notations:} Bold letters stand for vectors or matrices whereas a symbol not in bold font represents a scalar. $|.|$ and $\left\|.\right\|$ refer to the absolute value of a scalar and the 2-norm of a vector. $\mathcal{E}\left\{.\right\}$ refers to the averaging operator.

\vspace{-0.0cm}
\section{System Model}\label{system}
\begin{figure}
 \begin{minipage}[c]{.5\linewidth}
   \centerline{\includegraphics[width=0.9\columnwidth]{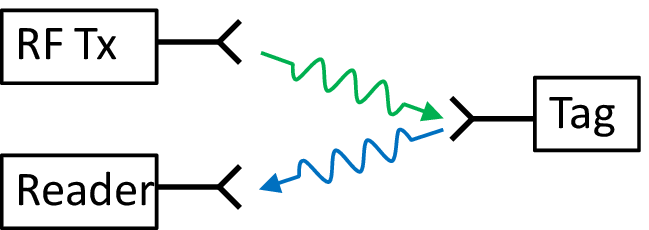}}
  \end{minipage}\hfill
 \begin{minipage}[c]{.5\linewidth}
   \centerline{\includegraphics[width=0.9\columnwidth]{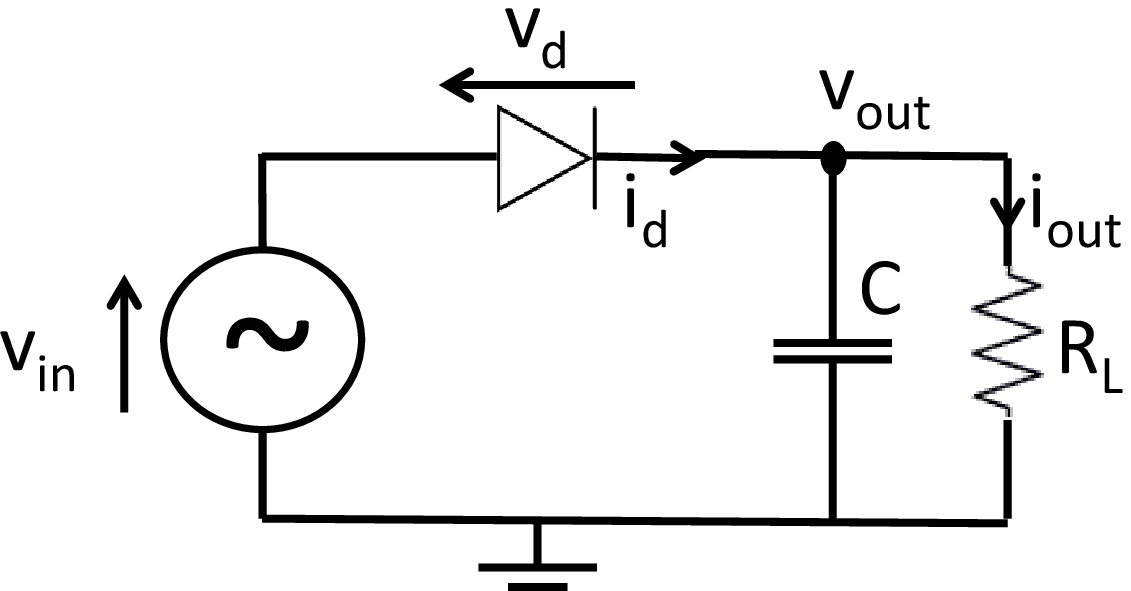}}
  \end{minipage}
  \caption{System architecture (left) and single diode rectifier at the tag (right).}
  \label{fig_sim_0}
  \vspace{-0.0cm}
\end{figure}

The overall system architecture is illustrated in Fig \ref{fig_sim_0} (left).
\vspace{-0.4cm}
\subsection{Received Signal at the Tag}
Consider a multisine signal (with $N$ sinewaves) transmitted by an RF transmitter at time $t$ over a single antenna
\begin{align}\label{WPT_waveform}
x(t)=\Re\left\{\sum_{n=0}^{N-1}w_{n}e^{j2\pi f_n t}\right\},
\end{align}
with $w_{n}=s_{n}e^{j\phi_{n}}$ where $s_{n}$ and $\phi_{n}$ refer to the amplitude and phase of the $n^{th}$ sinewave at frequency $f_n$, respectively. We assume for simplicity that the frequencies are evenly spaced, i.e.\ $f_n=f_0+n\Delta_f$ with $\Delta_f$ the frequency spacing. The magnitudes and phases of the sinewaves can be collected into vectors $\mathbf{s}$ and $\mathbf{\Phi}$. The $n^{th}$ entry of $\mathbf{s}$ and $\mathbf{\Phi}$ are written as $s_{n}$ and $\phi_{n}$, respectively. The transmitter is subject to a transmit power constraint $\mathcal{E}\big\{\left|x\right|^2\big\}=\frac{1}{2}\left\|\mathbf{s}\right\|_F^2\leq P$.

\par The transmit waveform propagates through a multipath channel and is received at the single-antenna tag as
\begin{align}
y(t)&=\sum_{n=0}^{N-1}s_{n}A_{n} \cos(2\pi f_n t+\psi_{n})\label{received_signal_ant_m}\\
&=\Re\left\{\sum_{n=0}^{N-1} h_n w_n e^{j 2\pi f_n t}\right\}
\end{align}
where $h_{n}=A_{n}e^{j \bar{\psi}_{n}}$ is the forward channel frequency response at frequency $f_n$. The amplitude $A_{n}$ and the phase $\psi_{n}$ are such that $A_{n}e^{j \psi_{n}}=A_{n}e^{j \left(\phi_{n}+\bar{\psi}_{n}\right)}=e^{j \phi_{n}}h_{n}$.

\vspace{-0.2cm}
\subsection{Tag's Operation}
We assume the tag only performs binary modulation. Binary 0 corresponds to a perfect impedance matching that completely absorbs the incoming signal (i.e. the reflection coefficient is 0). The signal absorbed by the tag during binary 0 operation is conveyed to a rectifier that converts the incoming RF signal into DC current. Binary 1 corresponds to a perfect impedance mismatch that competely reflects the incoming signal (i.e. the reflection coefficient is 1).  The signal reflected during binary 1 operation is backscattered to a reader, whose objective is to decide upon the sequence of transmitted bits (0 or 1). 

\vspace{-0.2cm}
\subsection{Rectenna Model and DC Current at the Tag}
We will assume the same rectenna model as in \cite{Clerckx:2015,Clerckx:2016b}. The rectenna is made of an antenna and a rectifier. The antenna model reflects the power transfer from the
antenna to the rectifier through the matching network. A lossless antenna
can be modelled as a voltage source $v_s(t)$ followed by a
series resistance $R_{ant}$. Let $Z_{in} = R_{in} + j X_{in}$ denote the
input impedance of the rectifier with the matching network.
Assuming perfect matching during binary operation 0 ($R_{in} = R_{ant}$, $X_{in} = 0$), all the
incoming RF power $P_{in,av}$ is transferred to the rectifier and
absorbed by $R_{in}$, so that $P_{in,av} = \mathcal{E}\big\{\left|v_{in}(t)\right|^2\big\}/R_{in}$ with $v_{in}(t)=v_{s}(t)/2$ the input voltage to the rectifier as per Fig \ref{fig_sim_0} (right). Since $P_{in,av} =\mathcal{E}\big\{\left|y(t)\right|^2\big\}$, $v_{in}(t)=y(t)\sqrt{R_{in}}=y(t)\sqrt{R_{ant}}$.

\par Consider a rectifier composed of a single diode followed by a low-pass filter with load ($R_L$). Denoting the voltage drop across the diode as $v_d(t)=v_{in}(t)-v_{out}(t)$ where  $v_{out}(t)$ is the output voltage across the load resistor (see Fig \ref{fig_sim_0}), a tractable behavioural diode model is obtained by Taylor series expansion of the diode characteristic equation $i_d(t)=i_s \big(e^{\frac{v_d(t)}{n v_t}}-1 \big)$ (with $i_s$ the reverse bias saturation current, $v_t$ the thermal voltage, $n$ the ideality factor equal to $1.05$) around a quiescent operating point $v_d=a$, namely
\begin{equation}\label{DiodeCurrent}
i_d(t)=\sum_{i=0}^{\infty }k_i' \left(v_d(t)-a\right)^i,
\end{equation}
where $k_0'=i_s\big(e^{\frac{a}{n v_t}}-1\big)$ and $k_i'=i_s\frac{e^{\frac{a}{n v_t}}}{i!\left(n v_t\right)^i}$, $i=1,\ldots,\infty$.

\par Assume a steady-state response and an ideal low pass filter such that $v_{out}(t)$ is at constant DC level. Choosing $a=\mathcal{E} \left\{ v_d(t) \right\}=-v_{out}$, \eqref{DiodeCurrent} can be simplified as $i_d(t)=\sum_{i=0}^{\infty }k_i' v_{in}(t)^i=\sum_{i=0}^{\infty }k_i' R_{ant}^{i/2} y(t)^i$. Truncating the expansion to order 4, the DC component of $i_{d}(t)$ is the time average of the diode current, and is obtained as $i_{out}\approx k_0'+k_2' R_{ant}\mathcal{E}\left\{y(t)^2\right\}+k_4' R_{ant}^2\mathcal{E}\left\{y(t)^4\right\}$.

\subsection{Backscatter Signal and SNR at the Reader}
The backscatter signal received at the reader is given by
\begin{align}\label{back_rec_signal}
z(t)
&=m\Re\left\{\sum_{n=0}^{N-1} h_{r,n} h_n w_n  e^{j 2\pi f_n t}\right\}+n(t)
\end{align}
where $m$ equals 0 or 1 for binary operation 0 and 1, respectively.
The quantity $n(t)$ is the AWGN and $h_{r,n}=A_{r,n}e^{j \bar{\psi}_{r,n}}$ is the frequency response of the backward channel (from tag to reader) on frequency $n$.

\par After applying a product detector to each frequency and assuming ideal low pass filtering, the baseband signal on each frequency $n$ is given by
\begin{equation}
z_n=h_{r,n} h_n w_n m +n_n
\end{equation} 
where $n_n \sim \mathcal{CN}(0,\sigma^2)$. The SNR after Maximum Ratio Combining (MRC) is finally given by
\begin{equation}
\rho\left(\mathbf{s}\right)=\frac{\sum_{n=0}^{N-1}\left|h_{r,n} h_n w_n\right|^2}{\sigma^2}=\frac{\sum_{n=0}^{N-1} A_{r,n}^2 A_n^2 s_n^2}{\sigma^2}.
\end{equation}

\subsection{CSIT Assumption}
We assume perfect CSIT, i.e.\ the forward $h_n$ and backscatter $h_n h_{r,n}$ channels are perfectly known $\forall n$ to the RF transmitter, so as to shape the transmit waveform dynamically as a function of the channel states to maximize $i_{out}$ and $\rho$. The backscatter channel $h_n h_{r,n}$ can be obtained at the RF transmitter by letting the reader send pilots, reaching the RF transmitter through backscattering. Backscatter and forward channels can then be estimated and obtained at the RF transmitter \cite{Yang:2015}.
\par We also assume that the concatenated channel $h_{r,n} h_n w_n$ is perfectly known to the reader to perform MRC.

\section{Waveform Optimization and SNR-Energy Region Characterization}\label{section_backscatter_waveform}

\par Subject to a transmit power constraint $\frac{1}{2}\left\|\mathbf{s}\right\|^2\leq P$ and under the assumption of perfect CSIT, the maximization of the SNR suggests an adaptive single-sinewave strategy (ASS) that consists in transmitting all power on a single sinewave, namely the one corresponding to the strongest channel $\bar{n} = \arg \max_i A_i A_{i,r}$. On the other hand, the maximization of the harvested energy, namely $i_{out}$, is shown in \cite{Clerckx:2016b} to be equivalent to maximizing the quantity 
\begin{equation}\label{diode_model_2}
z_{DC}\left(\mathbf{s},\mathbf{\Phi}\right)=k_2 R_{ant}\mathcal{E}\left\{y(t)^2\right\}+k_4 R_{ant}^2\mathcal{E}\left\{y(t)^4\right\}
\end{equation}
where $k_i=\frac{i_s}{i!\left(n v_t\right)^i}$, $i=2,4$\footnote{Assuming $i_s=5 \mu A$, a diode ideality factor $n=1.05$ and $v_t=25.86 mV$, typical values are given by $k_2=0.0034$ and $k_4=0.3829$.}. The maximization of \eqref{diode_model_2} suggests allocating power over multiple sinewaves, and those with stronger frequency-domain channel gains are allocated more power, in order to exploit the non-linearity of the rectifier and the frequency diversity \cite{Clerckx:2016b}. Hence the design of efficient waveforms for backscatter communication is subject to a tradeoff between maximizing received SNR at the reader and maximizing harvested energy at the tag. Characterizing this SNR-energy tradeoff and the corresponding waveform design is the objective of this section.

\begin{table*}
\begin{align}\label{z_DC}
z_{DC}(\mathbf{s},\mathbf{\Phi})&=\frac{k_{2}}{2}R_{ant}\left[\sum_{n=0}^{N-1} s_{n}^2A_{n}^2\right]+\frac{3k_{4}}{8}R_{ant}^2\left[\sum_{\mycom{n_0,n_1,n_2,n_3}{n_0+n_1=n_2+n_3}}\Bigg[\prod_{j=0}^3s_{n_j}A_{n_j}\Bigg]\cos(\psi_{n_0}+\psi_{n_1}-\psi_{n_2}-\psi_{n_3})\right].
\end{align}\hrulefill
\end{table*}

\par We can now define the achievable SNR-harvested energy (or more accurately SNR-DC current) region as
\begin{multline}
C_{SNR-I_{DC}}(P)\triangleq\Big\{(SNR,I_{DC}):SNR\leq \rho(\mathbf{s}),\Big. \\
\Big.I_{DC}\leq z_{DC}(\mathbf{s},\mathbf{\Phi}), \frac{1}{2}\left\|\mathbf{s}\right\|^2\leq P \Big\}.
\end{multline}
Optimal values $\mathbf{s}^{\star}$,$\mathbf{\Phi}^{\star}$ are to be found in order to enlarge as much as possible $C_{SNR-I_{DC}}$. The expression of $z_{DC}$ is provided in \eqref{z_DC} after plugging \eqref{received_signal_ant_m} into \eqref{diode_model_2}.


\par We note that the phases of the waveform $\phi_n$ influences $z_{DC}$ but not the SNR. Hence we can choose the phases as in point-to-point WPT in \cite{Clerckx:2015}, namely $\phi_{n}^{\star}=-\bar{\psi}_{n}$. This guarantees that all arguments of the cosine functions in $z_{DC}$ are equal to 0 in \eqref{z_DC}, which can simply be written as
\begin{multline}\label{z_DC_opt_phase}
z_{DC}(\mathbf{s},\mathbf{\Phi}^{\star})=\frac{k_{2}}{2}R_{ant}\left[\sum_{n=0}^{N-1} s_{n}^2A_{n}^2\right]\\
+\frac{3k_{4}}{8}R_{ant}^2\sum_{\mycom{n_0,n_1,n_2,n_3}{n_0+n_1=n_2+n_3}}\prod_{j=0}^3s_{n_j}A_{n_j}.
\end{multline}
$\mathbf{\Phi}^{\star}$ is obtained by collecting $\phi_{n}^{\star}$ $\forall n$ into a vector.  

\par Recall from \cite{Chiang:2007} that a monomial is defined as the function $g:\mathbb{R}_{++}^{N}\rightarrow\mathbb{R}:g(\mathbf{x})=c x_1^{a_1}x_2^{a_2}\ldots x_N^{a_N}$
where $c>0$ and $a_i\in\mathbb{R}$. A sum of $K$ monomials is called a posynomial and can be written as $f(\mathbf{x})=\sum_{k=1}^K g_k(\mathbf{x})$ with $g_k(\mathbf{x})=c_k x_1^{a_{1k}}x_2^{a_{2k}}\ldots x_N^{a_{Nk}}$ where $c_k>0$. As we can see from \eqref{z_DC_opt_phase}, $z_{DC}(\mathbf{s},\mathbf{\Phi}^{\star})$ is a posynomial. 

\par In order to identify the achievable SNR-energy region, we formulate the optimization problem as an energy maximization problem subject to transmit power and SNR constraints
\begin{align}\label{back_opt_problem}
\max_{\mathbf{s}} \hspace{0.3cm}&z_{DC}(\mathbf{s},\mathbf{\Phi}^{\star})\\
\textnormal{subject to} \hspace{0.3cm} &\frac{1}{2}\left\|\mathbf{s}\right\|^2\leq P,\\
& \rho(\mathbf{s})\geq \overline{SNR}.
\end{align}
It therefore consists in maximizing a posynomial subject to constraints. 
Unfortunately this problem is not a standard Geometric Program (GP) but it can be transformed to an equivalent problem by introducing an auxiliary variable $t_0$
\begin{align}\label{back_opt_problem_eq}
\min_{\mathbf{s},t_0} \hspace{0.3cm} &1/t_0\\
\textnormal{subject to} \hspace{0.3cm} &\frac{1}{2}\left\|\mathbf{s}\right\|^2\leq P,\\
&t_0/z_{DC}(\mathbf{s},\mathbf{\Phi}^{\star})\leq1,\\
&\frac{\overline{SNR}}{\rho(\mathbf{s})}\leq 1.\label{back_opt_problem_eq_4}
\end{align}
This is known as a Reversed Geometric Program. A similar problem also appeared in the WPT waveform optimization \cite{Clerckx:2015,Clerckx:2016b} and the rate-energy region characterization of Simultaneous Wireless Information and Power Transfer \cite{Clerckx:2016}. Note that $1/z_{DC}(\mathbf{s},\mathbf{\Phi}^{\star})$ and $1/\rho(\mathbf{s})$ are not posynomials, therefore preventing the use of standard GP tools. The idea is to replace the last two inequalities (in a conservative way) by making use of the arithmetic mean-geometric mean inequality.

\par Let $\left\{g_k(\mathbf{s},\mathbf{\Phi}^{\star})\right\}$ be the monomial terms in the posynomial $z_{DC}(\mathbf{s},\mathbf{\Phi}^{\star})=\sum_{k=1}^K g_k(\mathbf{s},\mathbf{\Phi}^{\star})$. Similarly we define $\left\{f_{n}(\mathbf{s})\right\}$ as the set of monomials of the posynomial $\rho(\mathbf{s})=\sum_{n=0}^{N-1}f_{n}(\mathbf{s})$ with $f_{n}(\mathbf{s})=s_n^2 A_n^2 A_{r,n}^2/\sigma^2$. For a given choice of $\left\{\gamma_k\right\}$ and $\left\{\beta_{n}\right\}$ with $\gamma_k,\beta_{n}\geq 0$ and $\sum_{k=1}^K \gamma_k=\sum_{i=1}^{I} \beta_{n}=1$, we perform single condensations and write the standard GP as
\begin{align}\label{standard_GP_back}
\min_{\mathbf{s},t_0} \hspace{0.3cm} &1/t_0\\
\textnormal{subject to} \hspace{0.3cm} &\frac{1}{2}\left\|\mathbf{s}\right\|^2\leq P,\\ 
&t_0\prod_{k=1}^K\left(\frac{g_k(\mathbf{s},\mathbf{\Phi}^{\star})}{\gamma_k}\right)^{-\gamma_k}\leq1,\\
&\overline{SNR}\prod_{n=0}^{N-1}\left(\frac{f_{n}(\mathbf{s})}{\beta_{n}}\right)^{-\beta_{n}}\leq 1.\label{standard_GP_back_3}
\end{align}
\par It is important to note that the choice of $\left\{\gamma_k,\beta_{n}\right\}$ plays a great role in the tightness of the AM-GM inequality. An iterative procedure can be used where at each iteration the standard GP \eqref{standard_GP_back}-\eqref{standard_GP_back_3} is solved for an updated set of $\left\{\gamma_k,\beta_{n}\right\}$. Assuming a feasible set of magnitude $\mathbf{s}^{(i-1)}$ at iteration $i-1$, compute at iteration $i$ $\gamma_k=g_k(\mathbf{s}^{(i-1)},\mathbf{\Phi}^{\star})/z_{DC}(\mathbf{s}^{(i-1)},\mathbf{\Phi}^{\star})$ $k=1,\ldots,K$ and $\beta_{n}= f_{n}(\mathbf{s}^{(i-1)})/\rho(\mathbf{s}^{(i-1)})$, $n=0,\ldots,N-1$, and then solve problem \eqref{standard_GP_back}-\eqref{standard_GP_back_3} to obtain $\mathbf{s}^{(i)}$. Repeat the iterations till convergence. The whole optimization procedure is summarized in Algorithm \ref{Algthm_OPT_back}. The successive approximation method used in the
Algorithm \ref{Algthm_OPT_back} is also known as a successive convex approximation. It cannot guarantee to converge to the global solution of the original problem, but yields a point fulfilling the KKT conditions \cite{Chiang:2007}.

\begin{algorithm}
\caption{Backscatter Communication Waveform}
\label{Algthm_OPT_back}
\begin{algorithmic}[1]
\State \textbf{Initialize}: $i\gets 0$, $\overline{SNR}$, $\mathbf{\Phi}^{\star}$, $\mathbf{s}$, $z_{DC}^{(0)}=0$
\label{Algthm_OPT_WIPT_step_initialize}
\Repeat
    \State $i\gets i+1$, $\ddot{\mathbf{s}}\gets \mathbf{s}$
    \State $\gamma_k\gets g_k(\ddot{\mathbf{s}},\mathbf{\Phi}^{\star})/z_{DC}(\ddot{\mathbf{s}},\mathbf{\Phi}^{\star})$, $k=1,\ldots,K$  
    \label{Algthm_OPT_back_step_gamma}
    \State $\beta_{n}\gets f_{n}(\ddot{\mathbf{s}})/\rho(\ddot{\mathbf{s}})$, $n=0,\ldots,N-1$  
    \label{Algthm_OPT_back_step_gamma_2}
    \State  $\mathbf{s} \gets \arg \min \eqref{standard_GP_back}-\eqref{standard_GP_back_3}$
    \label{Algthm_OPT_back_step_OPT}
    \State $z_{DC}^{(i)} \gets z_{DC}(\mathbf{s},\mathbf{\Phi}^{\star})$
\Until{$\left|z_{DC}^{(i)} - z_{DC}^{(i-1)} \right| < \epsilon$ \text{or} $i=i_{\max}$ }
\end{algorithmic}
\end{algorithm}

\vspace{-0.2cm}
\section{Simulation Results}\label{simulations}

\begin{figure}[!t]
\centering
\includegraphics[width=0.9\columnwidth]{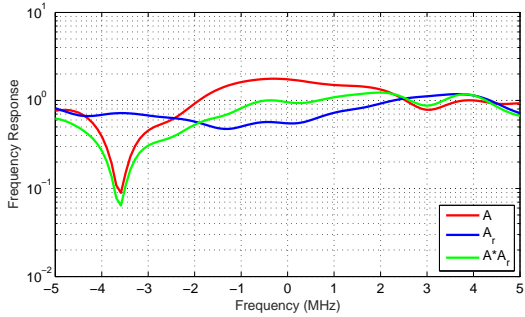}
\caption{Channel frequency responses over a 10 MHz bandwidth.}
\label{fig_sim_1}
\end{figure}

We consider a centre frequency of 5.18GHz, 36dBm EIRP, 2dBi receive and transmit antenna gain at the tag and 2dBi receive antenna gain at the reader. The path loss between the transmitter and the tag and between the tag and the reader is 58dB for each link. A NLOS channel power delay profile is obtained from model B \cite{Medbo:1998b}. The channel taps each with an average power $\beta_l$ are independent, circularly symmetric complex random Gaussian distributed and normalized such that $\sum_l \beta_l=1$. This leads to an average receive power of -20dBm at the tag and -74dBm at the reader. The noise power $\sigma^2$ at the reader is fixed to -84dB. The simulation is run over a channel realization with a bandwidth $B=1,10$ MHz. The frequency responses of the forward and backward channels are illustrated in Fig \ref{fig_sim_1}. The channel frequency response within the 1 MHz bandwisth is obtained by looking at Fig 1 between -0.5 MHz and 0.5 MHz. The frequency spacing of the multisine waveform is fixed as $\Delta_f=B/N$ and the $N$ sinewaves are centered around 5.18 GHz.

\begin{figure}[!t]
\centering
\includegraphics[width=0.9\columnwidth]{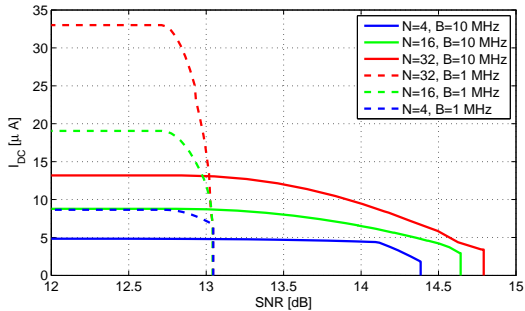}
\caption{SNR-$I_{DC}$ trade-off with B=1MHz and B=10MHz.}
\label{fig_sim_2}
\end{figure}

For the channel frequency responses of Fig \ref{fig_sim_1}, Algorithm \ref{Algthm_OPT_back} is used, along with CVX \cite{CVX}, to compute the optimal waveform and the corresponding SNR-$I_{DC}$ tradeoff, illustrated in Fig \ref{fig_sim_2} for B=1MHz and B=10MHz. The extreme point on the x-axis (SNR maximization) is achieved using the ASS strategy. On the other hand, the maximum energy is in general achieved by allocating transmit power over multiple subcarriers (as a consequence of the non-linearity of the rectifier) \cite{Clerckx:2016b}. A \textit{first} observation from Fig \ref{fig_sim_2} is that SNR and $I_{DC}$ are indeed subject to a fundamental tradeoff, i.e.\ increasing one of them is likely to result in a decrease of the other one. Nevertheless, as the channel becomes more frequency flat or the bandwidth decreases, the SNR-$I_{DC}$ appears more rectangular. A \textit{second} observation is that an increase in the number of frequency components $N$ of the multisine waveform results in an enlarged SNR-energy region. Indeed, by increasing $N$, the waveform exploits the nonlinearity of the rectifier and a frequency diversity gain, the latter being beneficial to both SNR and energy. A \textit{third} observation is that the shape of the SNR-energy region highly depends on the channel realizations and bandwidth. In particular, for the specific channel realization of Fig \ref{fig_sim_1}, we note that the 1MHz bandwidth favours higher $I_{DC}$ while the 10 MHz bandwidth favours higher SNR. This can be explained as follows. Recall first that $I_{DC}$ is a function of the forward channel amplitudes $A_n$ $\forall n$, while the SNR is a function of the backscatter channel $A_n A_{r,n}$. From Fig \ref{fig_sim_1}, $A_n$ reaches its peak for frequencies between -1 MHz and 0.5 MHz. Since the multisine waveform with a power allocation over multiple frequency components helps increasing $I_{DC}$, allocating the $N$ frequencies uniformly within the 1MHz bandwidth leads to higher $I_{DC}$ than that obtained with a 10MHz bandwidth (which exhibits deep fades). On the other hand, $A_n A_{r,n}$ exhibits its largest gain around 2MHz, which is outside the 1MHz bandwidth. Since ASS maximizes the SNR, larger SNRs are obtained on the 10MHz channel.  

\vspace{-0.2cm}
\section{Conclusions}\label{conclusions}
The paper derived a methodology to design adaptive transmit multisine waveforms for backscatter communications and characterize the fundamental tradeoff between conveying energy to the tag and enhancing the SNR of the backscatter communication link. Future interesting works consist in addressing the design of waveforms and the characterization of the SNR-energy region for more general setup including multiple antennas, multiple transmitters and multiple tags. The problem of CSI acquisition and its impact on the SNR-energy region is also of significant interest.

\ifCLASSOPTIONcaptionsoff
  \newpage
\fi

\end{document}